\documentstyle[openbib,12pt,fleqn]{article}
\renewcommand{\baselinestretch}{1.2}

\newcommand{\fg}[1]{\item {\label{#1}}}
\newcommand{\rf}[1]{~\cite{#1}}

\newcommand{\barr}{\begin{array}}
\newcommand{\bea}{\begin{eqnarray}}
\newcommand{\beq}{\begin{equation}}
\newcommand{\ear}{\end{array}}
\newcommand{\eea}{\end{eqnarray}}

\newcommand{\eeq}{\end{equation}}

\newcommand{\spao}[1]{\mbox{\hspace{#1}}}
\newcommand{\spav}[1]{\parbox{1mm}{\vspace*{#1}}}

\begin{document}
\bibliographystyle{unsrt}

$\ $   
\vskip 1.9 cm\noindent
{\LARGE\bf A New Determinant for }
{\LARGE\bf Quantum Chaos \\ }
\spav{3mm}\\
{\normalsize Predrag Cvitanovi\'c and Per E. Rosenqvist
  $\footnotemark\footnotetext
        {Contribution to the proceedings of
         the International School for Advanced Studies
         workshop ``From Classical to Quantum Chaos", Trieste, July 1992;
         {\em Nuovo Cimento}, to appear.
        }$
}\\
\baselineskip 5mm
{\normalsize Niels Bohr Institute,\\ Blegdamsvej 17, DK-2100 Copenhagen \O
         \\ }
\spav{4mm}\\
\spao{12mm}{\small\bf Abstract\\}
\spav{2mm}\\
\spao{16mm}
{\small\parbox{13cm}{\spao{4mm}
We propose a new type of approximation to quantum determinants,
``quantum Fredholm determinant", and conjecture that, compared
to the
quantum Selberg zeta functions derived from Gutzwiller
semiclassical trace formulas, such determinants
have a larger domain of analyticity
for Axiom A hyperbolic systems. The conjecture
is supported by a numerical investigation of the 3-disk repeller.
}\\}
\spav{4mm}
\setcounter{footnote}{0}

Dynamical zeta functions\rf{ruelle}, Fredholm determinants\rf{grot}
and quantum Selberg zeta functions\rf{gutbook,voros} have recently been
established as powerful tools for evaluation of classical and
quantum averages in low dimensional chaotic dynamical
systems\rf{AACI}~-\rf{CHAOS92}.
The convergence of cycle expansions\rf{cycprl}
of zeta functions and Fredholm
determinants depends on their analytic properties; particularly strong
results exist for nice (Axiom A) hyperbolic systems, for which the
dynamical zeta functions are holomorphic\rf{groth},
and the Fredholm determinants
are entire functions\rf{frie,Rugh92}. In this note, motivated by the
recent results of Eckhardt and Russberg\rf{ER92},
we conjecture that in contrast to the
quantum Selberg zeta function, for nice hyperbolic systems the {\em quantum}
Fredholm determinant (introduced below) is entire, {\em i.e} free of poles.

For 2-dimensional Hamiltonian systems
the dynamical zeta function\rf{ruelle} is given by
\beq
1/\zeta=\prod_p (1-t_p)
\,\, ,\quad
1/\zeta_k=\prod_p (1-t_p/\Lambda_p^{k})
         = \exp\left( - \sum_p \sum_{r=1}^\infty {1 \over r}
                      ({t_p / {\Lambda_p^k}})^r \right)
\,\, ,
\label{dynzeta}
\eeq
the Fredholm determinant\rf{AACI,Rugh92,losal} is given by
\beq
F=\prod_p\prod_{k=0}^{\infty} (1-t_p/\Lambda_p^{k})^{k+1}
=\prod_{k=0}^{\infty} 1/\zeta_k^{k+1}
\,\, ,
\label{2d_Fred}
\eeq
and the quantum Selberg zeta function\rf{voros} is given by
\beq
Z = \prod_p \prod_{k=0}^\infty
   \left( 1 -{ t_p / {\Lambda_p^k}}  \right)
=\prod_{k=0}^{\infty} 1/\zeta_k
\,\, .
\label{2d_quant}
\eeq

In the above, $t_p$
is a weight associated with the cycle $p$,
and the subscript $ p $
runs through all distinct prime cycles.
A prime cycle is a single traversal of the orbit;
its label is a non-repeating symbol string.
The cycle weight $t_p$ depends on the average evaluated.
Following refs. \cite{ER92,gasp,eck} we shall perform our
numerical tests on the 3-disk repeller. For such systems, the
cycle weight is given by\rf{CEflows}
\beq
t_p = z^{n_p} e^{s T_p}/ {|\Lambda_p|}
\,\, ,
\label{t_p_class}
\eeq
in the evaluation of escape rates and correlation spectra, and by
\beq
t_p = z^{n_p} { e^{-i S_p/\hbar + \nu_p}
         \over {\sqrt{|\Lambda_p|} }}
\,\, ,
\label{t_p_quant}
\eeq
in the evaluation of the semiclassical approximation\rf{voros,gutzp}
to quantum resonances. Here $T_p$ is the $p$-cycle period,
$S_p$ is its action, $\nu_p$ the Maslov index and $\Lambda_p$ the
expanding eigenvalue. $z$ is a bookkeeping variable that
keeps track of the topological cycle length $n_p$, used to
expand zeta functions and determinants:
\beq
F(z) = \sum_{k=0}^{\infty} C_{k} z^k
\,\, .
\label{cyc_exp}
\eeq
In calculations $z$ is set to $z=1$.

If the dynamical evolution can be cast in terms of a transfer
operator multiplicative along the flow, if the corresponding mapping
(for ex., return map for a Poincar\'e section of the flow) is
analytic, and if the topology of the repeller is given by a finite
Markov partition, then the Fredholm determinant (\ref{2d_Fred})
with classical weight (\ref{t_p_class}) is entire\rf{Rugh92}.
In this case the cycle expansion coefficients (\ref{cyc_exp})
fall off asymptotically faster than
exponentially\rf{Rugh92,losal},
as $C_n \approx \Lambda^{-n^{3/2}}$.
This estimate is in agreement with
numerical tests of ref.~\cite{Rugh92}, as well as our and
ref.~\cite{ER92} numerical results for the 3-disk repeller, see
fig.~1.
However, as it is not known how quickly the asymptotics should set in,
such numerical results can be misleading: for example, for a larger
disk-disk spacing, preasymptotic oscillations are visible in
fig.~2. (Such oscillations can be observed already in simple
1-dimensional repellers).

On the basis of close analogy between the classical and the quantum zeta
functions\rf{eck}, it has been hoped\rf{C92} that for
nice hyperbolic systems the quantum Selberg zeta functions (\ref{2d_quant})
should also be entire. However, it has not
been possible to extend the classical Fredholm determinant proof\rf{Rugh92}
to the quantum case, essentially because the composition of the
semiclassical propagators\rf{gutbook} is not multiplicative
along the classical trajectory, but requires additional saddlepoint
approximations. Indeed, Eckhardt and Russberg\rf{ER92} have
recently established by numerical studies that the 3-disk
quantum Selberg zeta functions have poles.

In refs.~\cite{CPR90,AACII}
heuristic arguments were developed
for 1-dimen\-si\-on\-al mappings to explain how the poles of individual
$1/\zeta_k$
cancel against the zeros of
$1/\zeta_{k+1}$, and thus
conspire to make the corresponding Fredholm determinant entire.
Eckhardt and Russberg have repeated this analysis for the
$1/\zeta_k$ terms in the quantum Selberg zeta function
(\ref{2d_quant}). They find numerically that $1/\zeta_0$ has a
double pole coinciding with the leading zero of $1/\zeta_1$.
Consequently $1/\zeta_0$, $1/\zeta_0 \zeta_1$ and $Z$ all have the
same leading pole, and coefficients in their cycle expansions fall off
exponentially with the same slope. Our numerical tests on the 3-disk
system, fig.~1, support this conclusion.

Why should $1/\zeta_0$ have a {\em double} leading pole? The
double pole is not as surprising as it might seem at the first glance;
indeed,
the theorem that establishes that
the classical Fredholm determinant (\ref{2d_Fred})
is entire implies that the
poles in $1/\zeta_k$ must have right multiplicities in order
that they be cancelled in the $ F = \prod 1/\zeta_k^{k+1}$ product.
More explicitely, $1/\zeta_k$ can be expressed in terms of
weighted Fredholm determinants
\beq
F_k =
    \exp\left( - \sum_p \sum_{r=1}^\infty {1 \over r}
                 { {({t_p / {\Lambda_p^k}})^r}
                  \over
                   {(1-1/\Lambda^r_p)^2}
                 }
        \right)
\label{F_k}
\eeq
($F_0 = F$ defined in (\ref{2d_Fred})) by inserting the
identity
\[
1= {1 \over {(1-1/\Lambda)^2}}
   -{2\over \Lambda} {1 \over {(1-1/\Lambda)^2}}
   + {1 \over {\Lambda^2}} {1 \over {(1-1/\Lambda)^2}}
\]
into the exponential representation (\ref{dynzeta}) of $1/\zeta_k$.
This yields
\beq
1/\zeta_k = { {F_k F_{k+2}} \over {F_{k+1}^2}}
\,\, ,
\label{doub_pole}
\eeq
and we conclude that for 2-dimensional Hamiltonian flows the
dynamical zeta function $1/\zeta_k$ has a {\em double} leading pole
coinciding with the leading zero of the $F_{k+1}$
Fredholm determinant. It is easy to check that the infinite
product $\prod 1/\zeta_k^{k+1}$ collapses to $F_0 = F$.
 $F_k$ can be interpreted as the Fredholm determinant
$\det(1-{\cal L}_k)$ of the weighted transfer operator
\beq
{\cal L}_k^t (y,x) =  \Lambda^t(x)^{-k} \phi^t(x)
                  \delta ( y - f^t(x))
\,\, ,
\label{weight_L}
\eeq
where $\Lambda^t(x)$ is the expanding eigenvalue of the
Jacobian transverse to the flow,
and $\phi^t(x)$ is any smooth weight multiplicative along the
trajectory.

The numerical results of ref.~\cite{ER92} suggest that the
{\em quantum Fredholm determinant}, {\em i.e.} the
Fredholm determinant (\ref{2d_Fred}) with the {\em quantum} weights
(\ref{t_p_quant}) may be entire, and that, in the spirit of the
thermodynamical formalism\rf{ruelle,sinai,bowen},
the quantum evolution operator should be approximated by a
classical transfer operator with a quantum weighting factor:
\beq
{\cal L}^t (y,x) =  \sqrt{|\Lambda^t(x)|} e^{-i S^t(x)/\hbar + \nu_p}
                  \delta ( y - f^t(x))
\,\, .
\label{quant_L}
\eeq
The difference between the two infinite products
(\ref{2d_Fred}) and (\ref{2d_quant}) can be traced to the
quantum $1/\sqrt{\det(1-J_p)}$ weight that
arises in the saddle point expansion derivation of
the Gutzwiller trace formula; the delta-function transfer
operator (\ref{quant_L}) leads to the cycle weight
$1/|\det(1-J_p)|$ instead.
Both the quantum Fredholm determinant and the quantum
zeta function yield  the same leading zeros, given by $1/\zeta_0$.
They presumably differ in nonleading zeros (with larger
imaginary part of the complex energy), but as
the quantum Selberg zeta function (\ref{2d_quant}) is the leading term of a
semiclassical approximation, with the size of corrections unknown,
the physical significance of these nonleading zeros remains unclear.

The transfer operator
(\ref{quant_L}) is problematic as it stands,
because it is not multiplicative
along the trajectory. While for the Jacobians $J_{ab}=J_a J_b$ for
two successive segments $a$ and $b$ along the trajectory, the
corresponding expanding eigenvalues are {\em not} multiplicative,
$\Lambda_{ab} \neq \Lambda_a \Lambda_b$, and consequently
(\ref{quant_L}) does not satisfy the assumptions required by
the theorem of ref.~\cite{Rugh92}. Nevertheless, such transfer operators
have been routinely used in, for ex., evaluation of partial
dimensions\rf{par_dims}. Our numerical results support the conjecture
that $\Lambda^k$ weighted determinant has enlarged domain of convergence.

In conclusion, we have proposed and tested numerically
a new approximation to the
quantum determinant, and conjectured that it has better
analyticity properties than the commonly used quantum Selberg zeta
function. The quantum Fredholm determinant suggests a
starting approximation to the quantum propagator different from
the usual used Van Vleck semiclassical propagator.
The new determinant is expected to be of practical utility
as for nice hyperbolic systems its convergence is superior to
that of the quantum Selberg zeta functions.

\spav{9mm}
{\bf Acknowledgements:}
The authors are grateful to
B. Eckhardt and G. Russberg for kindly communicating
their results prior to publication.
\\\\

\renewcommand{\baselinestretch} {1}

{\bf {Figure captions}}
\vskip 20pt

\begin{enumerate}

\fg{fig1}
$\log_{10} | C_n |$, the contribution of cycles of length $n$
to the cycle expansion $\sum C_n z^n$ for
$A_1$ symmetric subspace resonance for 3-disk repeller
with center spacing - disk radius ratio $R:a= 3:1$,
evaluated at the lowest resonance, wave number
$k=7.8727 - 0.3847\,i $. Shown are:
($\circ$) $1/\zeta_0$,
($\nabla$) the quantum Selberg zeta function,
($\Box$) $1/\zeta_0 \zeta_1^2$,
and
($\triangle$) the quantum Fredholm determinant.
Exponential falloff implies that $1/\zeta_0$ and the quantum Selberg zeta
have the same leading pole, cancelled in the $1/\zeta_0 \zeta_1^2$
product.
For comparison,
($\Diamond$) the classical Fredholm determinant coefficients are plotted
as well; cycle expansions for both Fredholm determinants appear
to follow the asymptotic estimate $C_n \approx \Lambda^{-n^{3/2}}$.

\fg{fig2}  Same as fig.~1, but with $ R : a = 6 : 1 $.
This illustrates possible pitfalls of numerical tests of asymptotics;
the quantum Fredholm determinant appears to have the same pole as
the quantum $1/\zeta_0 \zeta_1^2$, but
there is we have no estimate on the size of preasymptotic oscillations in
cycle expansions, it is difficult to draw reliable conclusions from
such numerics.

\end{enumerate}

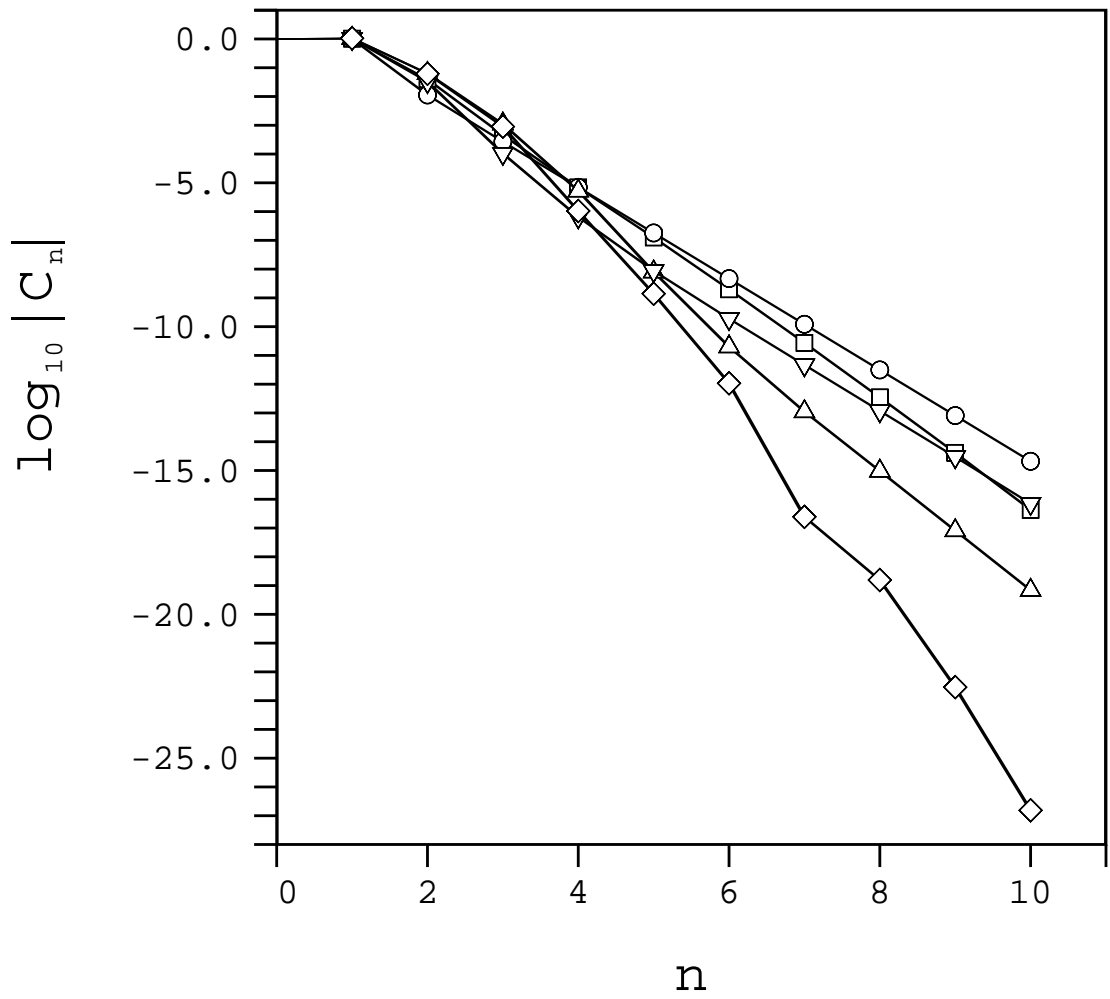
\end{document}